\documentclass[preprint]{ptephy} %%% choose this for arXiv

\preprintnumber{KYUSHU-HET-177}

\usepackage{amssymb}
\usepackage{amsthm}
\usepackage{amsmath}
\usepackage{booktabs}
\usepackage{bbm}
\usepackage{mathtools}
\DeclareMathOperator{\tr}{tr}
\DeclareMathOperator{\Tr}{Tr}

\numberwithin{equation}{section}

\usepackage{hyperref}

\begin{document}

\title{One-loop perturbative coupling of $A$ and~$A_\star$ through the chiral
overlap operator}

\author{%
\name{\fname{Hiroki} \surname{Makino}}{1},
\name{\fname{Okuto} \surname{Morikawa}}{1},
and
\name{\fname{Hiroshi} \surname{Suzuki}}{1,\ast}}

\address{%
\affil{1}{Department of Physics, Kyushu University
744 Motooka, Nishi-ku, Fukuoka, 819-0395, Japan}
\email{hsuzuki@phys.kyushu-u.ac.jp}}

\begin{abstract}
We study the one-loop effective action defined by the chiral overlap operator
in the four-dimensional lattice formulation of chiral gauge theories by
Grabowska and Kaplan. In the tree-level continuum limit, the left-handed
component of the fermion is coupled only to the original gauge field~$A$, while
the right-handed one is coupled only to~$A_\star$, which is given by the
gradient flow of~$A$ with infinite flow time. In this paper, we show that the
continuum limit of the one-loop effective action contains local interaction
terms between $A$ and~$A_\star$, which do not generally vanish even if the
gauge representation of the fermion is anomaly free. We argue that the presence
of such interaction terms can be regarded as undesired gauge symmetry-breaking
effects in the formulation.
\end{abstract}
\subjectindex{B01, B05, B31, B57}
\maketitle

\section{Introduction and discussion}
\label{sec:1}
Recently, Grabowska and Kaplan constructed a four-dimensional lattice
formulation of chiral gauge theories~\cite{Grabowska:2016bis}, starting from
their five-dimensional domain-wall formulation
in~Ref.~\cite{Grabowska:2015qpk}.\footnote{A closely related six-dimensional
domain-wall formulation is given in~Ref.~\cite{Fukaya:2016ofi}.} A salient
feature of this formulation is that the lattice Dirac operator depends on two
gauge fields: one is the original gauge field~$A$, and the other is~$A_\star$,
which is given by the gradient flow~\cite{Narayanan:2006rf,Luscher:2009eq,%
Luscher:2010iy,Luscher:2011bx} of~$A$ with infinite flow time. In the
tree-level continuum limit of the formulation, the left-handed component of
the fermion is coupled only to~$A$, while the right-handed one (called the
fluffy mirror fermion or ``fluff'') is coupled only to~$A_\star$. Up to a
subtlety associated with the topological charge~\cite{Grabowska:2015qpk,%
Grabowska:2016bis,Okumura:2016dsr,Makino:2016auf}, $A_\star$ basically becomes
pure gauge after the infinite-time flow. Then this setup would be regarded as
the system of the left-handed Weyl fermion coupled to the gauge field~$A$ (in
the spirit of~Ref.~\cite{AlvarezGaume:1983cs}). Since the flow equation
preserves the gauge covariance~\cite{Narayanan:2006rf,Luscher:2009eq,%
Luscher:2010iy,Luscher:2011bx}, $A_\star$ transforms gauge covariantly under
the gauge transformation. Then the fermion determinant is manifestly gauge
invariant in this formulation, \emph{even if the gauge representation is
anomalous}. It is crucial to understand, therefore, how this formulation
\emph{fails\/} when the gauge representation is anomalous. It is conceivable
that the locality plays a key role for this but no definite argument has been
given yet.

So far, the explicit form of the four-dimensional lattice Dirac operator in the
above formulation has been obtained only when the transition from~$A$
to~$A_\star$ along the flow is ``abrupt'' or ``sudden''; the resulting Dirac
operator is referred to as the \emph{chiral overlap operator\/} in the present
paper and is denoted by~$\Hat{\mathcal{D}}_\chi$. As noted above, in the
tree-level continuum limit~\cite{Grabowska:2016bis},
\begin{equation}
   am\Hat{\mathcal{D}}_\chi\stackrel{a\to0}{\to}
   \gamma_\mu D_\mu(A)P_-+\gamma_\mu D_\mu(A_\star)P_+,
\label{eq:(1.1)}
\end{equation}
where $a$ is the lattice spacing, $m$ is a parameter of mass dimension one,
$D_\mu(A)$ ($D_\mu(A_\star)$) is the covariant derivative defined with~$A$
($A_\star$), $\gamma_\mu$ is the Dirac matrix, and~$P_\pm=(1\pm\gamma_5)/2$
are the chirality projection operators. Thus, the lattice Dirac operator does
not produce any coupling between two gauge fields, $A$ and~$A_\star$, in the
tree-level approximation.

In this paper, we investigate how the above situation is modified under
radiative corrections. We thus study the fermion one-loop effective action
defined by
\begin{equation}
   \ln\mathcal{Z}[A,A_\star]
   \equiv\ln\int\prod_x\left[d\psi(x)d\Bar{\psi}(x)\right]
   \exp\left[-a^4\sum_x\Bar{\psi}(x)\Hat{\mathcal{D}}_\chi\psi(x)\right],
\label{eq:(1.2)}
\end{equation}
where the two gauge fields $A$ and~$A_\star$ are regarded as independent
non-dynamical variables. In the present paper, we assume that the gauge field
is perturbative and the Dirac operator has no normalizable zero modes. What we
will show in this paper is
\begin{equation}
   \delta\delta_\star\ln\mathcal{Z}[A,A_\star]
   =-\int d^4x\,\mathcal{L}(A,A_\star;\delta A,\delta_\star A_\star)
\label{eq:(1.3)}
\end{equation}
in the continuum limit, where
$\mathcal{L}(A,A_\star;\delta A,\delta_\star A_\star)$ is a \emph{local\/}
polynomial of its arguments and their spacetime derivatives. In this expression
and in what follows, the infinitesimal variation~$\delta$ acts only on~$A$ but
not on~$A_\star$,
\begin{equation}
   \delta A\neq0,\qquad\delta A_\star\equiv0,
\label{eq:(1.4)}
\end{equation}
while $\delta_\star$ acts in an opposite way,
\begin{equation}
   \delta_\star A\equiv0,\qquad\delta_\star A_\star\neq0.
\label{eq:(1.5)}
\end{equation}
Equation~\eqref{eq:(1.3)} tells us that through fermion one-loop diagrams, two
gauge fields $A$ and~$A_\star$ acquire local couplings. The locality
of~$\mathcal{L}(A,A_\star;\delta A,\delta_\star A_\star)$ is expected because in
the lattice Dirac operator~$\Hat{\mathcal{D}}_\chi$ the coupling between $A$
and~$A_\star$ is~$O(a)$ and thus the coupling emerges only through ultraviolet
divergences. We will further find that
$\mathcal{L}(A,A_\star;\delta A,\delta_\star A_\star)$ does not vanish even if
the gauge representation of the fermion is anomaly free.

What is a possible implication of our observation~\eqref{eq:(1.3)}? To find
this, let us classify the terms in $\ln\mathcal{Z}[A,A_\star]$ according to
their dependences as,
\begin{equation}
   \ln\mathcal{Z}[A,A_\star]
   ={\mit\Gamma}_0[A]+{\mit\Gamma}_1[A,A_\star]+{\mit\Gamma}_2[A_\star],
\label{eq:(1.6)}
\end{equation}
where ${\mit\Gamma}_1[A,A_\star]$ consists of cross terms between $A$
and~$A_\star$.\footnote{We note that ${\mit\Gamma}_1[A,A_\star]$ is a local
functional of~$A$ and~$A_\star$ because it can be reconstructed
from~$\mathcal{L}(A,A_\star;\delta A,\delta_\star A_\star)$ in an algebraic way.
On the other hand, ${\mit\Gamma}_0[A]$ and~${\mit\Gamma}_2[A_\star]$ are
non-local functionals of the argument.} Equation~\eqref{eq:(1.3)} thus shows
that
\begin{equation}
   \delta\delta_\star{\mit\Gamma}_1[A,A_\star]
   =-\int d^4x\,\mathcal{L}(A,A_\star;\delta A,\delta_\star A_\star)
\label{eq:(1.7)}
\end{equation}
in the continuum limit. On the other hand, by construction~\eqref{eq:(1.2)},
the effective action is invariant under the gauge transformation, if we
gauge-transform \emph{both\/} $A$ and~$A_\star$~\cite{Grabowska:2016bis}.
Expressing the \emph{gauge variations\/} by a superscript as
\begin{align}
   \delta^\omega A_\mu(x)&\equiv\partial_\mu\omega(x)+[A_\mu(x),\omega(x)],&
   \delta^\omega A_{\star\mu}(x)&=0,
\label{eq:(1.8)}
\\
   \delta_\star^\omega A_{\star\mu}(x)
   &\equiv\partial_\mu\omega(x)+[A_{\star\mu}(x),\omega(x)],&
   \delta_\star^\omega A_\mu(x)&=0,
\label{eq:(1.9)}
\end{align}
the gauge invariance implies
\begin{equation}
   (\delta^\omega+\delta_\star^\omega)\ln\mathcal{Z}[A,A_\star]=0
   \Rightarrow
   \delta(\delta^\omega+\delta_\star^\omega)\ln\mathcal{Z}[A,A_\star]=0.
\label{eq:(1.10)}
\end{equation}
Then, using Eqs.~\eqref{eq:(1.6)} and~\eqref{eq:(1.7)} in this relation, we
have
\begin{equation}
   \delta\delta^\omega{\mit\Gamma}_0[A]
   +\delta\delta^\omega{\mit\Gamma}_1[A,A_\star]
   =-\delta\delta_\star^\omega{\mit\Gamma}_1[A,A_\star]
   =\int d^4x\,\mathcal{L}(A,A_\star;\delta A,\delta_\star^\omega A_\star).
\label{eq:(1.11)}
\end{equation}

Now, the gauge field~$A_\star$ is given by the gradient flow of~$A$ for
infinite flow time. Thus, let us assume that $A_\star$ is pure gauge. Although
there exists a subtlety as to whether this is actually the case or not for
topologically non-trivial gauge field configurations~\cite{Grabowska:2016bis,%
Okumura:2016dsr}, this will certainly be the case for topologically trivial
configurations. Under this assumption, since the lattice gauge action~$S_G[A]$
(such as the plaquette action) with which the gauge field~$A$ is integrated
over will be gauge invariant, we may take a particular gauge in which
$A_\star\equiv0$.\footnote{We are grateful to Yoshio Kikukawa for pointing out
the simplicity occurring in this $A_\star=0$ gauge.} In this $A_\star=0$ gauge,
since ${\mit\Gamma}_1[A,A_\star]$ and~${\mit\Gamma}_2[A_\star]$
in~Eq.~\eqref{eq:(1.6)} are constants, from Eqs.~\eqref{eq:(1.6)}
and~\eqref{eq:(1.11)}, we have
\begin{equation}
   \delta\delta^\omega\ln\mathcal{Z}[A,0]
   =\int d^4x\,
   \mathcal{L}(A,A_\star=0;\delta A,
   \delta_\star^\omega A_\star|_{A_\star=0}).
\label{eq:(1.12)}
\end{equation}
We will see that the right-hand side does not vanish even if the gauge
representation is anomaly free. Thus, generally, configurations of the gauge
field $A$ (in a topologically trivial sector) are integrated with the sum of
the gauge-invariant action~$S_G[A]$ and gauge \emph{non-invariant\/} effective
action~$\ln\mathcal{Z}[A,0]$.\footnote{The Faddeev--Popov ghost term associated
with this $A_\star=0$ gauge would be
$S_{c\Bar{c}}=-2\int d^4x\,\tr[\Bar{c}_\mu(x)\partial_\mu c(x)]$ in terms of the
continuum theory. Since this does not contain the gauge field, it does not
influence our argument.} For example, we will see that
$\mathcal{L}(A,A_\star=0;\delta A,\delta_\star^\omega A_\star|_{A_\star=0})$
contains a term corresponding to the mass term of the gauge field. Such
gauge-breaking effects which are not related to the gauge anomaly should be
able to be removed by local counterterms. This expectation is explicitly
confirmed in~Appendix~\ref{sec:A}. Nevertheless, such a necessity for
counterterms to restore the gauge symmetry will be undesirable for a
possible non-perturbative formulation of chiral gauge theories. This is the
implication of our observation~\eqref{eq:(1.3)}. It appears that the
formulation of~Ref.~\cite{Grabowska:2016bis} with the chiral overlap
operator~$\Hat{\mathcal{D}}_\chi$ (i.e., the sudden flow case) should be
improved in some possible way. In the rest of this paper, we will explain how
Eq.~\eqref{eq:(1.7)} is obtained.

\section{Computation of Eq.~\eqref{eq:(1.7)}}
\label{sec:2}
\subsection{Basic formulation}
\label{sec:2.1}
The explicit form of the chiral overlap operator~$\Hat{\mathcal{D}}_\chi$ is
given by~\cite{Grabowska:2016bis},
\begin{equation}
   a\Hat{\mathcal{D}}_\chi
   =1+\gamma_5\left[1-(1-\epsilon_\star)\frac{1}{\epsilon\epsilon_\star+1}
   (1-\epsilon)\right],
\label{eq:(2.1)}
\end{equation}
where $\epsilon$ and~$\epsilon_\star$ are the sign
functions~\cite{Neuberger:1997fp,Neuberger:1998wv}
\begin{equation}
   \epsilon\equiv\frac{H_w(A)}{\sqrt{H_w(A)^2}},\qquad
   \epsilon_\star\equiv\frac{H_w(A_\star)}{\sqrt{H_w(A_\star)^2}}
\label{eq:(2.2)}
\end{equation}
of the Hermitian Wilson--Dirac operator
\begin{equation}
   H_w=\gamma_5\left[\frac{1}{2}\gamma_\mu(\nabla_\mu+\nabla_\mu^*)
   -\frac{1}{2}a\nabla_\mu\nabla_\mu^*-m\right],
\label{eq:(2.3)}
\end{equation}
where $\nabla_\mu$ and~$\nabla_\mu^*$ are forward and backward gauge covariant
lattice derivatives, respectively. The parameter $m$ is taken as~$0<am<2$.
From Eq.~\eqref{eq:(2.2)}, $\epsilon$ depends only on the gauge field~$A$ and
$\epsilon_\star$ only on~$A_\star$. By construction,
\begin{equation}
   \epsilon^2=\epsilon_\star^2=1.
\label{eq:(2.4)}
\end{equation}
Using these, one can confirm that
\begin{equation}
   \left[1-(1-\epsilon_\star)\frac{1}{\epsilon\epsilon_\star+1}
   (1-\epsilon)\right]^2=1
\label{eq:(2.5)}
\end{equation}
and, consequently, $\Hat{\mathcal{D}}_\chi$ in~Eq.~\eqref{eq:(2.1)} satisfies
the Ginsparg--Wilson relation~\cite{Ginsparg:1981bj}
\begin{equation}
   \gamma_5\Hat{\mathcal{D}}_\chi+\Hat{\mathcal{D}}_\chi\gamma_5
   =a\Hat{\mathcal{D}}_\chi\gamma_5\Hat{\mathcal{D}}_\chi.
\label{eq:(2.6)}
\end{equation}
It is then natural to introduce a modified
$\gamma_5$~\cite{Luscher:1998pqa,Niedermayer:1998bi},
\begin{equation}
   \Hat{\gamma}_5\equiv\gamma_5(1-a\Hat{\mathcal{D}}_\chi),
\label{eq:(2.7)}
\end{equation}
which satisfies
\begin{equation}
   (\Hat{\gamma}_5)^2=1,\qquad
   \Hat{\mathcal{D}}_\chi\Hat{\gamma}_5=-\gamma_5\Hat{\mathcal{D}}_\chi.
\label{eq:(2.8)}
\end{equation}
Note, however, that $\Hat{\gamma}_5$ is not Hermitian in the present
formulation, $\Hat{\gamma}_5^\dagger\neq\Hat{\gamma}_5$. From the first relation
of~Eq.~\eqref{eq:(2.8)}, one can define modified chiral projection operators
by
\begin{equation}
   \Hat{P}_\pm\equiv\frac{1}{2}(1\pm\Hat{\gamma}_5).
\label{eq:(2.9)}
\end{equation}
The chiral components of the fermion can then be defined as
\begin{align}
   \Hat{P}_-\psi_L(x)&=\psi_L(x),&\Bar{\psi}_L(x)P_+=\Bar{\psi}_L(x),
\label{eq:(2.10)}
\\
   \Hat{P}_+\psi_R(x)&=\psi_R(x),&\Bar{\psi}_R(x)P_-=\Bar{\psi}_R(x).
\label{eq:(2.11)}
\end{align}
Thanks to the second relation of~Eq.~\eqref{eq:(2.8)}, the action is
completely decomposed into the left-handed and right-handed components as
\begin{equation}
   a^4\sum_x\Bar{\psi}(x)\Hat{\mathcal{D}}_\chi\psi(x)
   =a^4\sum_x\left[\Bar{\psi}_L(x)\Hat{\mathcal{D}}_\chi\psi_L(x)
   +\Bar{\psi}_R(x)\Hat{\mathcal{D}}_\chi\psi_R(x)\right].
\label{eq:(2.12)}
\end{equation}

\subsection{Gauge currents and partial decoupling of the right-handed fluff
fermion}
\label{sec:2.2}
In the present paper, we assume that the gauge field is perturbative and the
Dirac operator has no normalizable zero modes in infinite volume. Then the
change of the effective action~\eqref{eq:(1.2)} under the variation of the
gauge field $\delta$~\eqref{eq:(1.4)}, for example, is given by
\begin{align}
   \delta\ln\mathcal{Z}[A,A_\star]
   &=-a^4\sum_x\left\langle\Bar{\psi}(x)
   \delta\Hat{\mathcal{D}}_\chi\psi(x)\right\rangle
\notag\\
   &=\Tr\delta\Hat{\mathcal{D}}_\chi
   \frac{1}{\Hat{\mathcal{D}}_\chi},
\label{eq:(2.13)}
\end{align}
where $\Tr\equiv\sum_x\tr$ and $\tr$ stands for the trace over the spinor
and gauge indices. In deriving this, we have used the fermion propagator,
\begin{equation}
   \left\langle\psi(x)\Bar{\psi}(y)\right\rangle
   =\frac{1}{\Hat{\mathcal{D}}_\chi}\frac{1}{a^4}\delta_{x,y}.
\label{eq:(2.14)}
\end{equation}
We will refer to~Eq.~\eqref{eq:(2.13)} (and a similar expression for the
variation~$\delta_\star$~\eqref{eq:(1.5)}) as the ``gauge current.'' Because
of~Eq.~\eqref{eq:(2.8)}, we may decompose the gauge current~\eqref{eq:(2.13)}
into two parts by inserting chiral projectors:
\begin{equation}
   \Tr\delta\Hat{\mathcal{D}}_\chi\frac{1}{\Hat{\mathcal{D}}_\chi}
   =\Tr\delta\Hat{\mathcal{D}}_\chi\Hat{P}_-
   \frac{1}{\Hat{\mathcal{D}}_\chi}P_+
   +\Tr\delta\Hat{\mathcal{D}}_\chi\Hat{P}_+
   \frac{1}{\Hat{\mathcal{D}}_\chi}P_-.
\label{eq:(2.15)}
\end{equation}
In the right-hand side of this expression, the first term can be regarded as a
collection of one-loop diagrams of the physical left-handed fermion containing
at least one interaction vertex with~$A$. Similarly, the second term can be
regarded as a collection of similar one-loop diagrams but of the right-handed
fluff fermion.

Interestingly, the last term of~Eq.~\eqref{eq:(2.15)}
\emph{identically vanishes\/} even with finite lattice spacings. This might be
regarded as a (partial) decoupling of the fluff fermions from the physical
gauge field~$A$ in the one-loop level; this is certainly a desired property. To
see this, we first note that because of~$\epsilon^2=\epsilon_\star^2=1$ the
chiral overlap operator~\eqref{eq:(2.1)} can be written as
\begin{equation}
   a\Hat{\mathcal{D}}_\chi
   =1+\gamma_5\left[1-2(1-\epsilon_\star)\frac{1}{\epsilon\epsilon_\star+1}
   \right].
\label{eq:(2.16)}
\end{equation}
Then since $\delta$~\eqref{eq:(1.4)} does not change $\epsilon_\star$,
noting again that $\epsilon^2=\epsilon_\star^2=1$ (and thus
$\delta\epsilon\epsilon=-\epsilon\delta\epsilon$), we have the following
sequence of equalities:
\begin{align}
   \delta a\Hat{\mathcal{D}}_\chi
   &=2\gamma_5(1-\epsilon_\star)\frac{1}{\epsilon\epsilon_\star+1}
   \delta\epsilon\epsilon_\star\frac{1}{\epsilon\epsilon_\star+1}
\notag\\
   &=\gamma_5(1-\epsilon_\star)(1-\epsilon_\star)
   \frac{1}{\epsilon\epsilon_\star+1}
   \delta\epsilon\epsilon_\star\frac{1}{\epsilon\epsilon_\star+1}
\notag\\
   &=\gamma_5(1-\epsilon_\star)
   \frac{1}{\epsilon\epsilon_\star+1}(1-\epsilon)
   \delta\epsilon\epsilon_\star\frac{1}{\epsilon\epsilon_\star+1}
\notag\\
   &=\gamma_5(1-\epsilon_\star)
   \frac{1}{\epsilon\epsilon_\star+1}
   \delta\epsilon(1+\epsilon)\epsilon_\star\frac{1}{\epsilon\epsilon_\star+1}
\notag\\
   &=\gamma_5(1-\epsilon_\star)
   \frac{1}{\epsilon\epsilon_\star+1}
   \delta\epsilon\epsilon_\star\frac{1}{\epsilon\epsilon_\star+1}
   (1+\epsilon_\star).
\label{eq:(2.17)}
\end{align}
On the other hand, from~Eqs.~\eqref{eq:(2.9)}, \eqref{eq:(2.7)},
and~\eqref{eq:(2.1)}, we have
\begin{equation}
   \Hat{P}_+=\frac{1+\Hat{\gamma}_5}{2}
   =\frac{1}{2}(1-\epsilon_\star)\frac{1}{\epsilon\epsilon_\star+1}(1-\epsilon).
\label{eq:(2.18)}
\end{equation}
These show that
\begin{equation}
   \delta a\Hat{\mathcal{D}}_\chi\Hat{P}_+=0
\label{eq:(2.19)}
\end{equation}
and thus the last term of~Eq.~\eqref{eq:(2.15)} identically vanishes. That is,
\begin{equation}
   \Tr\delta\Hat{\mathcal{D}}_\chi\frac{1}{\Hat{\mathcal{D}}_\chi}
   =\Tr\delta\Hat{\mathcal{D}}_\chi\Hat{P}_-
   \frac{1}{\Hat{\mathcal{D}}_\chi}P_+,\qquad
   \Tr\delta\Hat{\mathcal{D}}_\chi\Hat{P}_+
   \frac{1}{\Hat{\mathcal{D}}_\chi}P_-=0.
\label{eq:(2.20)}
\end{equation}
As relations being dual to these, we also have
\begin{equation}
   \Tr\delta_\star\Hat{\mathcal{D}}_\chi\frac{1}{\Hat{\mathcal{D}}_\chi}
   =\Tr\delta_\star\Hat{\mathcal{D}}_\chi\Hat{P}_+
   \frac{1}{\Hat{\mathcal{D}}_\chi}P_-,\qquad
   \Tr\delta_\star\Hat{\mathcal{D}}_\chi\Hat{P}_-
   \frac{1}{\Hat{\mathcal{D}}_\chi}P_+=0.
\label{eq:(2.21)}
\end{equation}

\subsection{Functional curl}
\label{sec:2.3}
The structure of the ``gauge current''~\eqref{eq:(2.20)} is quite analogous to
the covariantly regularized gauge current of the left-handed Weyl
fermion~\cite{Fujikawa:1983bg}, which leads to the covariant gauge
anomaly~\cite{Bardeen:1984pm}. This definition of the gauge current preserves
the gauge covariance even for anomalous cases at the expense of the Bose
symmetry in fermion one-loop diagrams. The Bose symmetry is restored (in the
continuum theory) if the gauge representation of the Weyl fermion is
anomaly free. The breaking of the Bose symmetry can be characterized by the
``functional curl''; this notion appears in various places in consideration of
the anomaly---see for example, Ref.~\cite{Banerjee:1986bu}
and~Sect.~6.6 of~Ref.~\cite{Fujikawa:2004cx}. In our present problem, the
analogue of the functional curl (associated with the right-handed fluff
fermion) would be
\begin{equation}
   \delta_\star\Tr\delta\Hat{\mathcal{D}}_\chi\Hat{P}_+
   \frac{1}{\Hat{\mathcal{D}}_\chi}P_-
   -\delta\Tr\delta_\star\Hat{\mathcal{D}}_\chi\Hat{P}_+
   \frac{1}{\Hat{\mathcal{D}}_\chi}P_-.
\label{eq:(2.22)}
\end{equation}
We expect that in the continuum limit this combination becomes local because
if we neglect the subtlety associated with the definition of the gauge current
in quantum theory (such as the covariant versus consistent), then the gauge
current would always be given by the derivative of the effective action and
then the combination such as~Eq.~\eqref{eq:(2.22)} would vanish. We will
shortly see that this expectation is correct. Note that
Eqs.~\eqref{eq:(2.20)}, \eqref{eq:(2.21)} and~\eqref{eq:(2.13)}
with~$\delta\to\delta_\star$ imply that
\begin{align}
   \delta_\star\Tr\delta\Hat{\mathcal{D}}_\chi\Hat{P}_+
   \frac{1}{\Hat{\mathcal{D}}_\chi}P_-
   -\delta\Tr\delta_\star\Hat{\mathcal{D}}_\chi\Hat{P}_+
   \frac{1}{\Hat{\mathcal{D}}_\chi}P_-
   &=-\delta\delta_\star\ln\mathcal{Z}[A,A_\star]
\notag\\
   &\stackrel{a\to0}{\to}\int d^4x\,
   \mathcal{L}(A,A_\star;\delta A,\delta_\star A_\star),
\label{eq:(2.23)}
\end{align}
where in the last equality we have used the notation in~Eq.~\eqref{eq:(1.3)};
thus the local functional~$\mathcal{L}(A,A_\star;\delta A,\delta_\star A_\star)$
in~Eq.~\eqref{eq:(1.3)} is given by (the integrand of) the functional
curl~\eqref{eq:(2.22)}.

\subsection{Functional curl~\eqref{eq:(2.22)} is a local functional}
\label{sec:2.4}
The following argument is almost identical to the one given
in~Ref.~\cite{Suzuki:1999qw} which tries to interpret the lattice formulation
of~Ref.~\cite{Luscher:1998du} in terms of the covariant gauge current. Instead
of Eq.~\eqref{eq:(2.22)} itself, it is convenient to consider
\begin{align}
   &\Delta_1\Tr\Delta_2\Hat{\mathcal{D}}_\chi\Hat{P}_+
   \frac{1}{\Hat{\mathcal{D}}_\chi}P_-
   -(1\leftrightarrow2)
\notag\\
   &=\Delta_1\Tr\Delta_2\Hat{\mathcal{D}}_\chi
   \frac{1}{\Hat{\mathcal{D}}_\chi}P_-
   -(1\leftrightarrow2)
\notag\\
   &=-\Tr\Delta_2\Hat{\mathcal{D}}_\chi
   \frac{1}{\Hat{\mathcal{D}}_\chi}
   \Delta_1\Hat{\mathcal{D}}_\chi
   \frac{1}{\Hat{\mathcal{D}}_\chi}P_-
   -(1\leftrightarrow2)
\notag\\
   &=\frac{1}{2}\Tr\Delta_2\Hat{\mathcal{D}}_\chi
   \frac{1}{\Hat{\mathcal{D}}_\chi}
   \Delta_1\Hat{\mathcal{D}}_\chi
   \frac{1}{\Hat{\mathcal{D}}_\chi}\gamma_5
   -(1\leftrightarrow2),
\label{eq:(2.24)}
\end{align}
where $\Delta\equiv\delta+\delta_\star$ stands for a general infinitesimal
variation of the gauge fields $A$ and~$A_\star$; in the very final step, we
will set $\Delta_1=\delta_\star$ and~$\Delta_2=\delta$.

Introducing the notation
\begin{equation}
   H\equiv\gamma_5\Hat{\mathcal{D}}_\chi,
\label{eq:(2.25)}
\end{equation}
the Ginsparg--Wilson relation~\eqref{eq:(2.6)} yields
\begin{align}
   &\frac{1}{H}\gamma_5+\Hat{\gamma}_5\frac{1}{H}=0,
\label{eq:(2.26)}
\\
   &\frac{1}{H}\Hat{\gamma}_5+\Hat{\gamma}_5\frac{1}{H}=-a,
\label{eq:(2.27)}
\\
   &\Hat{\gamma}_5\Delta H+\Delta H\Hat{\gamma}_5=0.
\label{eq:(2.28)}
\end{align}
 From the last two relations, we have
\begin{equation}
   \frac{1}{H}\Delta H\Hat{\gamma}_5
   =\Hat{\gamma}_5\frac{1}{H}\Delta H+a\Delta H.
\label{eq:(2.29)}
\end{equation}

Now Eq.~\eqref{eq:(2.24)} can be written as
\begin{equation}
   -\frac{1}{2}\Tr\Delta_1H\frac{1}{H}\Delta_2H\frac{1}{H}
   \gamma_5
   -(1\leftrightarrow2)
   =\frac{1}{2}\Tr\Delta_1H\frac{1}{H}\Delta_2H\Hat{\gamma}_5
   \frac{1}{H}
   -(1\leftrightarrow2),
\label{eq:(2.30)}
\end{equation}
where we have used~Eq.~\eqref{eq:(2.26)}. Then using Eq.~\eqref{eq:(2.29)},
\begin{equation}
   \frac{1}{2}\Tr\Delta_1H\frac{1}{H}\Delta_2H\Hat{\gamma}_5
   \frac{1}{H}
   -(1\leftrightarrow2)
   =\frac{1}{2}a\Tr\Delta_1H\Delta_2H\frac{1}{H}.
\label{eq:(2.31)}
\end{equation}
We then put $\Hat\gamma_5^2=1$ in the last expression and move one
$\Hat{\gamma}_5$ within the $\Tr$. Using Eqs.~\eqref{eq:(2.29)}
and~\eqref{eq:(2.28)}, we have
\begin{align}
   \frac{1}{2}a\Tr\Delta_1H\Delta_2H\frac{1}{H}
   &=-\frac{1}{2}a\Tr\Delta_1H\Delta_2H\frac{1}{H}
   -\frac{1}{2}a^2\Tr\Delta_1H\Delta_2H\Hat{\gamma}_5
\notag\\
   &=-\frac{1}{4}a^2\Tr\Delta_1H\Delta_2H\Hat{\gamma}_5
\notag\\
   &=-\frac{1}{8}a^2\Tr\Hat{\gamma}_5[\Delta_1H,\Delta_2H].
\notag\\
   &=-\frac{1}{8}\Tr\Hat{\gamma}_5
   [\Delta_1\Hat{\gamma}_5,\Delta_2\Hat{\gamma}_5].
\label{eq:(2.32)}
\end{align}

Finally, setting $\Delta_1=\delta_\star$ and~$\Delta_2=\delta$, for the
functional curl~\eqref{eq:(2.22)},
\begin{equation}
   \delta_\star\Tr\delta\Hat{\mathcal{D}}_\chi\Hat{P}_+
   \frac{1}{\Hat{\mathcal{D}}_\chi}P_-
   -\delta\Tr\delta_\star\Hat{\mathcal{D}}_\chi\Hat{P}_+
   \frac{1}{\Hat{\mathcal{D}}_\chi}P_-
   =-\frac{1}{8}\Tr\Hat{\gamma}_5
   [\delta_\star\Hat{\gamma}_5,\delta\Hat{\gamma}_5].
\label{eq:(2.33)}
\end{equation}
Since no inverse of the Dirac operator is involved in the right-hand side, the
functional curl is manifestly a local functional of $A$ and~$A_\star$.

We further rewrite Eq.~\eqref{eq:(2.33)} as follows: First, we note that
\begin{equation}
   -\frac{1}{8}\Tr\Hat{\gamma}_5
   [\delta_\star\Hat{\gamma}_5,\delta\Hat{\gamma}_5]
   =\frac{1}{4}\Tr\Hat{P}_+
   [\delta\Hat{\gamma}_5,\delta_\star\Hat{\gamma}_5].
\label{eq:(2.34)}
\end{equation}
As we have seen in~Eq.~\eqref{eq:(2.17)},
\begin{align}
   \delta\Hat{\gamma}_5
   &=-(1-\epsilon_\star)\frac{1}{\epsilon\epsilon_\star+1}
   \delta\epsilon\epsilon_\star
   \frac{1}{\epsilon\epsilon_\star+1}(1+\epsilon_\star),
\label{eq:(2.35)}
\\
   \delta_\star\Hat{\gamma}_5
   &=-(1+\epsilon)\frac{1}{\epsilon\epsilon_\star+1}
   \epsilon\delta_\star\epsilon_\star
   \frac{1}{\epsilon\epsilon_\star+1}(1-\epsilon),
\label{eq:(2.36)}
\end{align}
and using Eq.~\eqref{eq:(2.18)} and relations such as
$(1-\epsilon)(1-\epsilon_\star)=(\epsilon\epsilon_\star+1)(1-\epsilon_\star)$
and~$(1+\epsilon_\star)(1+\epsilon)=(\epsilon+\epsilon_\star)(1+\epsilon)$,
after some calculation we find that
\begin{equation}
   -\frac{1}{8}\Tr\Hat{\gamma}_5
   [\delta_\star\Hat{\gamma}_5,\delta\Hat{\gamma}_5]
   =-\frac{1}{2}\Tr(1-\epsilon_\star)
   \frac{1}{\epsilon+\epsilon_\star}
   \delta\epsilon\frac{1}{\epsilon+\epsilon_\star}\delta_\star\epsilon_\star.
\label{eq:(2.37)}
\end{equation}
We decompose this according to the number of~$\gamma_5$. Then the parity-odd
part of the functional curl~\eqref{eq:(2.22)} is given by
\begin{align}
   (\text{parity-odd part})
   &=\frac{1}{2}\Tr\epsilon_\star
   \frac{1}{\epsilon+\epsilon_\star}\delta\epsilon
   \frac{1}{\epsilon+\epsilon_\star}\delta_\star\epsilon_\star
\notag\\
   &=\frac{1}{4}\Tr\epsilon_\star
   \frac{1}{\epsilon+\epsilon_\star}\delta\epsilon
   \frac{1}{\epsilon+\epsilon_\star}\delta_\star\epsilon_\star
   -\frac{1}{4}\Tr\epsilon
   \frac{1}{\epsilon+\epsilon_\star}\delta_\star\epsilon_\star
   \frac{1}{\epsilon+\epsilon_\star}\delta\epsilon,
\label{eq:(2.38)}
\end{align}
and the parity-even part is
\begin{align}
   (\text{parity-even part})
   &=-\frac{1}{2}\Tr
   \frac{1}{\epsilon+\epsilon_\star}\delta\epsilon
   \frac{1}{\epsilon+\epsilon_\star}\delta_\star\epsilon_\star
\notag\\
   &=-\frac{1}{4}\Tr
   \frac{1}{\epsilon+\epsilon_\star}\delta\epsilon
   \frac{1}{\epsilon+\epsilon_\star}\delta_\star\epsilon_\star
   -\frac{1}{4}
   \Tr\frac{1}{\epsilon+\epsilon_\star}\delta_\star\epsilon_\star
   \frac{1}{\epsilon+\epsilon_\star}\delta\epsilon.
\label{eq:(2.39)}
\end{align}
We see that the parity-odd part is anti-symmetric under the exchange
$A\leftrightarrow A_\star$, while the parity-even part is symmetric. We will
now present the continuum limit of these expressions.

\subsection{Continuum limit}
\label{sec:(2.5)}
The computational strategy for the continuum limit of Eqs.~\eqref{eq:(2.38)}
and~\eqref{eq:(2.39)} is identical to that of~Ref.~\cite{Makino:2016auf}. We
thus omit the details of the (very tedious) calculation and show only the
results. In what follows, we use the notation
\begin{align}
   C_\mu&\equiv A_{\star\mu}-A_\mu,
\label{eq:(2.40)}
\\
   \Bar{A}_\mu&\equiv\frac{1}{2}(A_\mu+A_{\star\mu}),
\label{eq:(2.41)}
\\
   \Bar{D}_\mu&\equiv\partial_\mu+[\Bar{A}_\mu,\cdot],
\label{eq:(2.42)}
\end{align}
and
\begin{align}
   F_{\mu\nu}&=\partial_\mu A_\nu-\partial_\nu A_\mu+[A_\mu,A_\nu],
\label{eq:(2.43)}
\\
   F_{\star\mu\nu}&=\partial_\mu A_{\star\nu}-\partial_\nu A_{\star\mu}
   +[A_{\star\mu},A_{\star\nu}],
\label{eq:(2.44)}
\\
   \Bar{F}_{\mu\nu}&=\partial_\mu\Bar{A}_\nu-\partial_\nu\Bar{A}_\mu
   +[\Bar{A}_\mu,\Bar{A}_\nu]
   =\frac{1}{2}F_{\mu\nu}+\frac{1}{2}F_{\star\mu\nu}
   -\frac{1}{4}[C_\mu,C_\nu].
\label{eq:(2.45)}
\end{align}

\begin{figure}[ht]
\begin{minipage}{0.45\columnwidth}
\begin{center}
\includegraphics[width=\columnwidth]{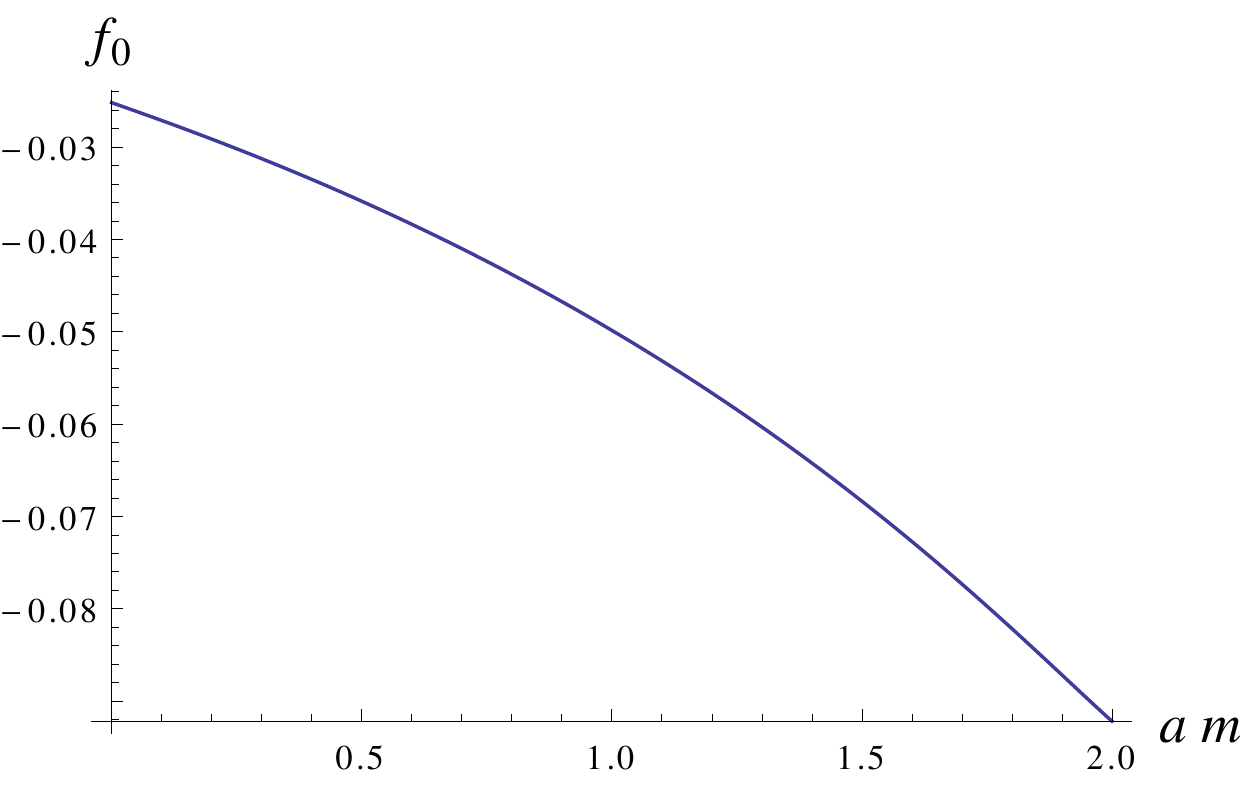}
\caption{$f_0(am)$.}
\label{fig:1}
\end{center}
\end{minipage}\hspace{3em}
\begin{minipage}{0.45\columnwidth}
\begin{center}
\includegraphics[width=\columnwidth]{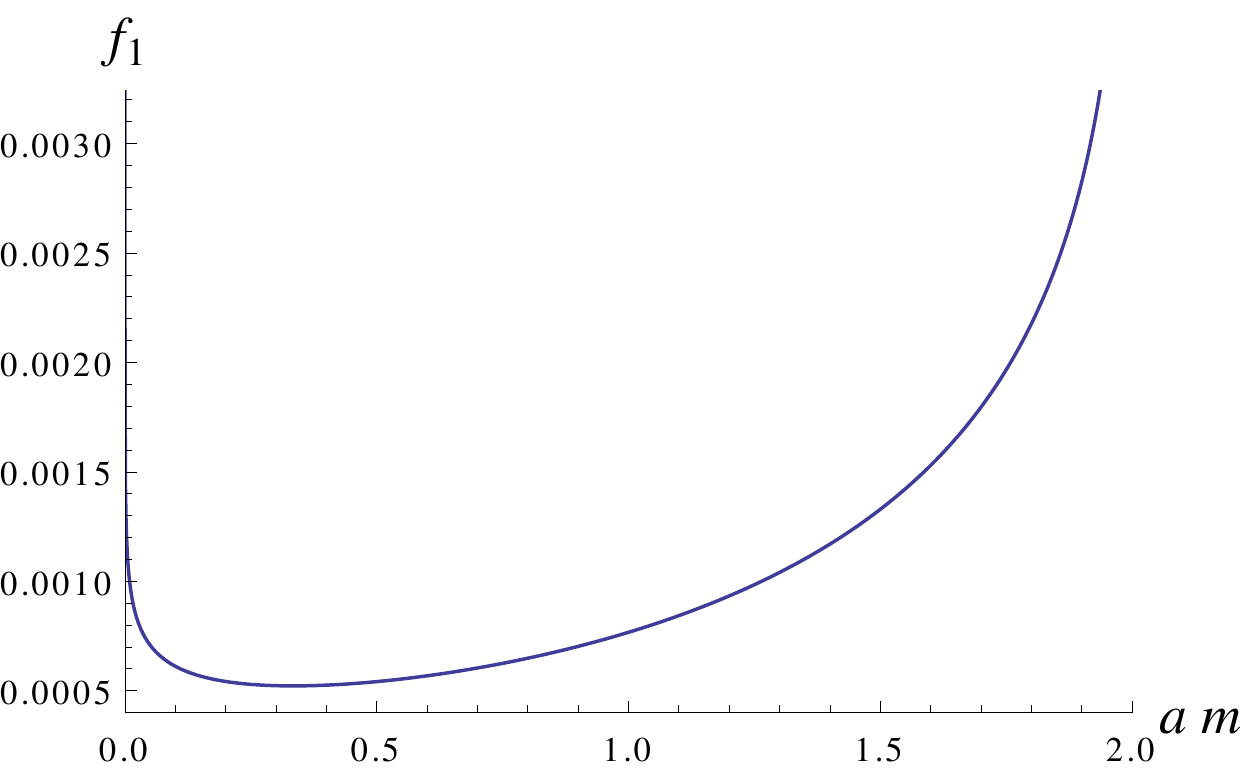}
\caption{$f_1(am)$.}
\end{center}
\end{minipage}
\end{figure}

\begin{figure}[ht]
\begin{minipage}{0.45\columnwidth}
\begin{center}
\includegraphics[width=\columnwidth]{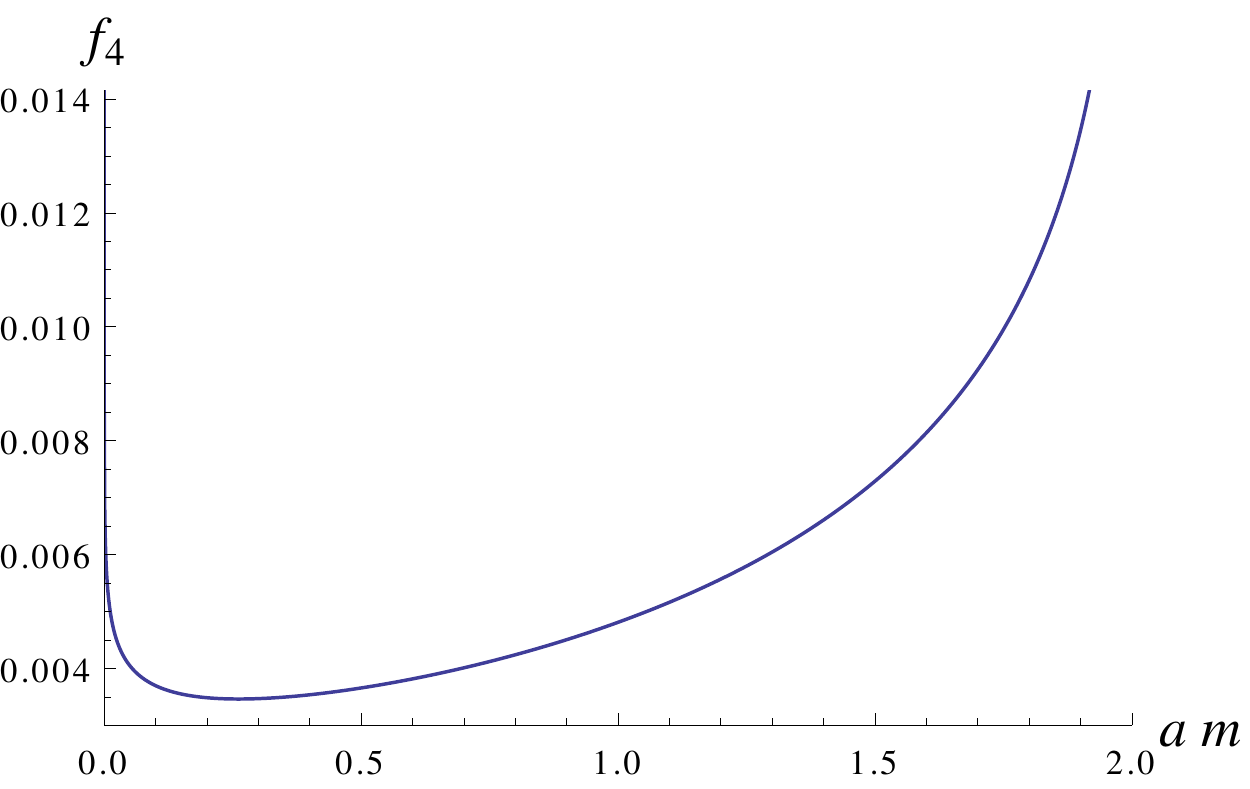}
\caption{$f_4(am)$.}
\end{center}
\end{minipage}\hspace{3em}
\begin{minipage}{0.45\columnwidth}
\begin{center}
\includegraphics[width=\columnwidth]{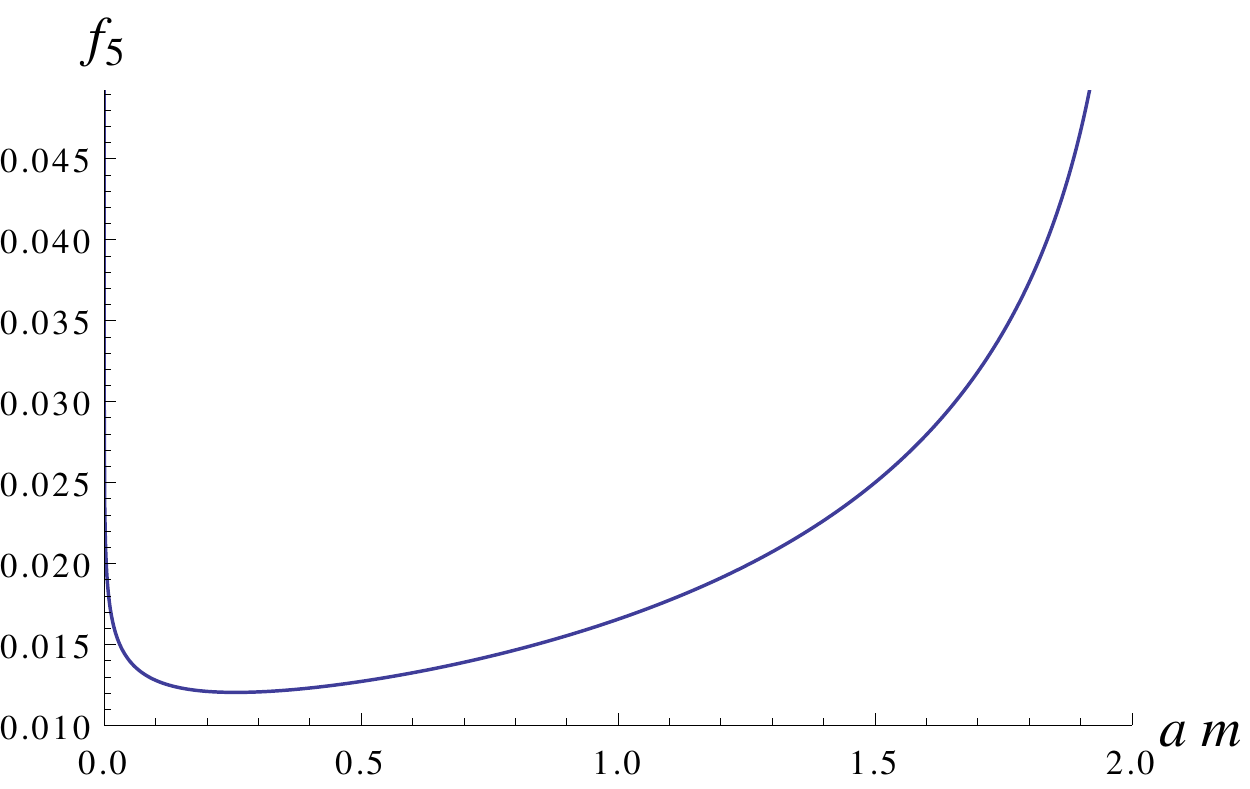}
\caption{$f_5(am)$.}
\label{fig:6}
\end{center}
\end{minipage}
\end{figure}

We also define the following lattice integrals. With the abbreviations,
\begin{align}
   s_\rho&\equiv\sin p_\rho,&c_\rho&\equiv\cos p_\rho,
\label{eq:(2.46)}
\\
   c&\equiv\sum_\mu(c_\mu-1)+am,&t&\equiv\sum_\mu s_\mu^2+c^2,
\label{eq:(2.47)}
\\
   \int_p&\equiv\int_{-\pi}^\pi\frac{d^4p}{(2\pi)^4},&&
\label{eq:(2.48)}
\end{align}
we define
\begin{align}
   f_0(am)
   &\equiv
   \int_p\left(-\frac{1}{4t}-\frac{s_\rho^2}{4t}-\frac{c c_\rho}{4t}\right),
\label{eq:(2.49)}
\\
   f_1(am)
   &\equiv\int_p\left(\frac{1}{64t^2}-\frac{c_\rho c_\sigma}{128t}
   +\frac{s_\rho^2s_\sigma^2}{32t^2}\right),
\label{eq:(2.50)}
\\
   f_2(am)
   &\equiv
   \int_p\left(-\frac{c_\rho c_\sigma}{32t}+\frac{7s_\rho^2 s_\sigma^2}{64t^2}
   +\frac{cs_\rho^2c_\sigma}{32t^2}+\frac{c^2c_\rho c_\sigma}{64t^2}\right),
\label{eq:(2.51)}
\\
   f_3(am)
   &\equiv
   \int_p\left(-\frac{c_\rho c_\sigma}{32t}+\frac{3s_\rho^2s_\sigma^2}{32t^2}
   -\frac{s_\rho^2}{32t^2}-\frac{cc_\rho}{32t^2}\right),
\label{eq:(2.52)}
\\
   f_4(am)
   &\equiv
   \int_p\left(\frac{1}{96t}+\frac{s_\rho^2}{96t}
   +\frac{cc_\rho}{96t}+\frac{1}{16t^2}\right),
\label{eq:(2.53)}
\\
   f_5(am)
   &\equiv
   \int_p\left(\frac{1}{16t}+\frac{c_\rho c_\sigma}{32t}
   +\frac{7}{32t^2}-\frac{c^2}{32t^2}+\frac{c c_\rho}{16t^2}
   +\frac{s_\rho^2}{32t^2}\right).
\label{eq:(2.54)}
\end{align}
The values of these lattice integrals as functions of the parameter~$am$
are depicted in~Figs.~\ref{fig:1}--\ref{fig:6}.

\begin{figure}[ht]
\begin{minipage}{0.45\columnwidth}
\begin{center}
\includegraphics[width=\columnwidth]{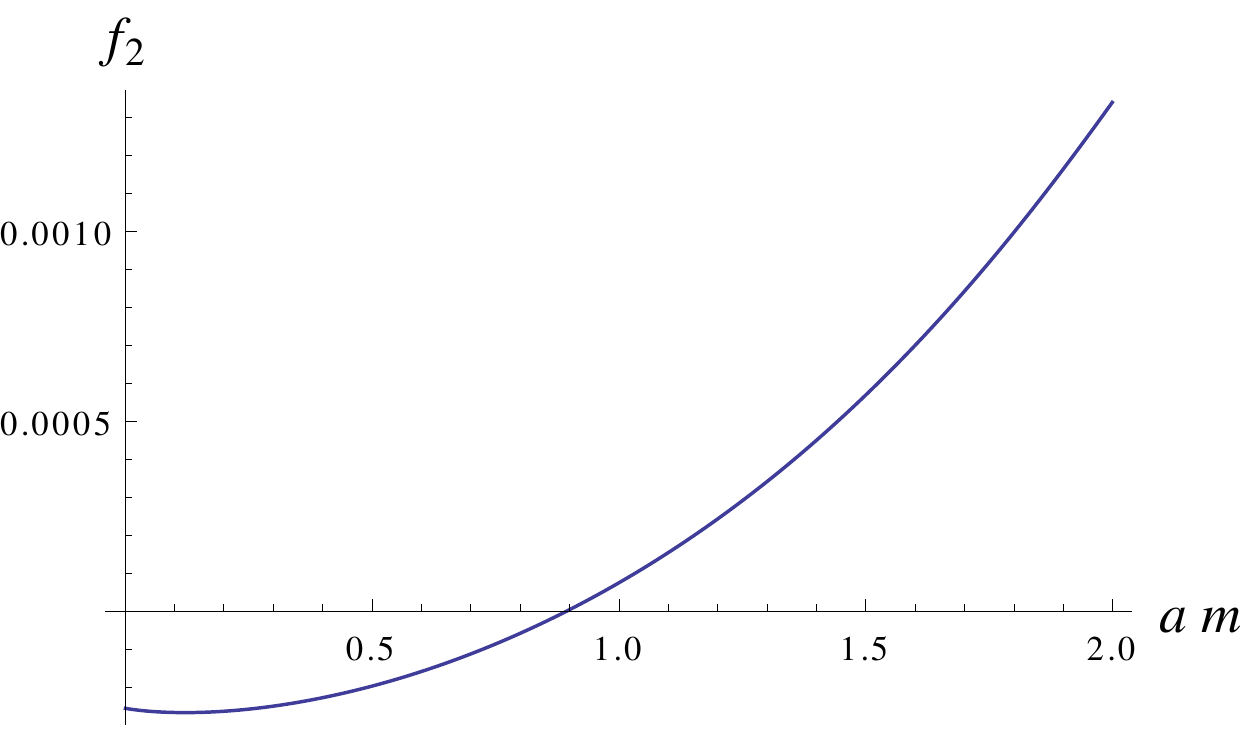}
\caption{$f_2(am)$.}
\end{center}
\end{minipage}\hspace{3em}
\begin{minipage}{0.45\columnwidth}
\begin{center}
\includegraphics[width=\columnwidth]{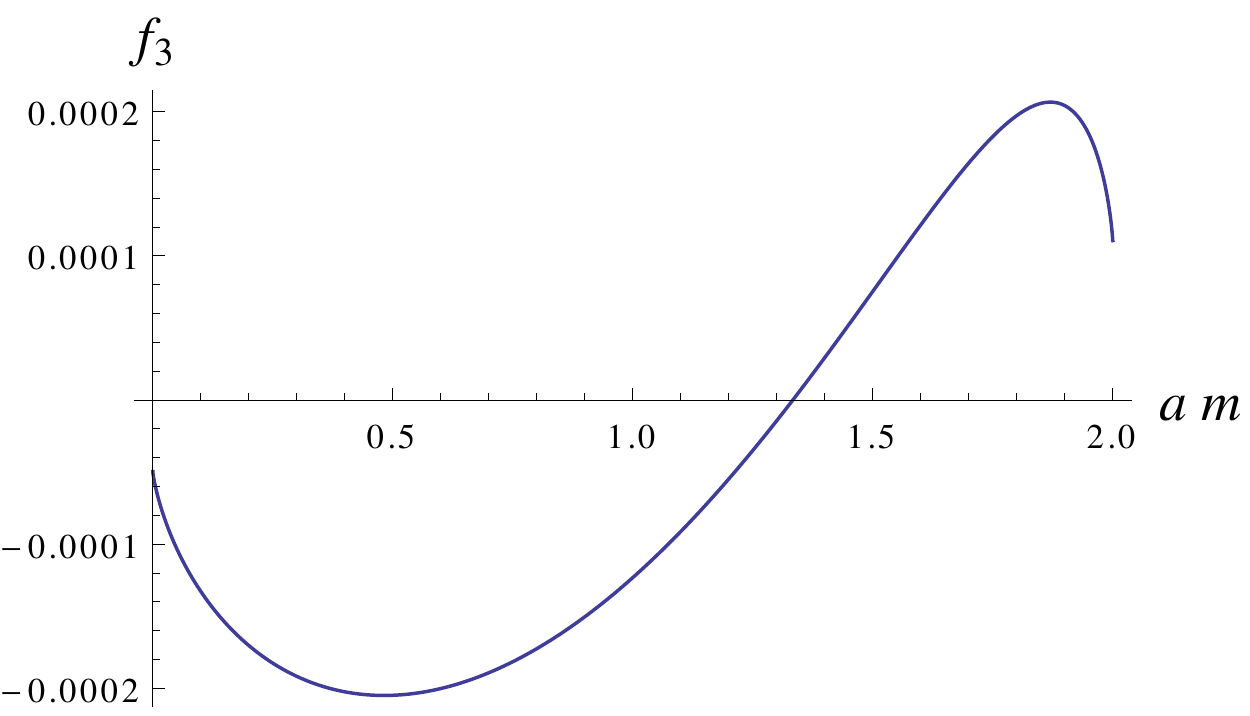}
\caption{$f_3(am)$.}
\end{center}
\end{minipage}
\end{figure}

The continuum limit of~Eq.~\eqref{eq:(2.38)}, i.e., the parity-odd part
of~$\mathcal{L}(A,A_\star;\delta A,\delta_\star A_\star)$,
recalling~Eq.~\eqref{eq:(2.23)}, is given by (omitting the symbol~$\tr$),
\begin{align}
   &\left.
   \mathcal{L}(A,A_\star;\delta A,\delta_\star A_\star)
   \right|_{\text{parity-odd}}
\notag\\
   &=-\frac{1}{32\pi^2}\epsilon_{\mu\nu\rho\sigma}
   \biggl[\left(\Bar{F}_{\mu\nu}+\frac{1}{12}[C_\mu,C_\nu]\right)
   \{\delta A_\rho,\delta_\star A_{\star\sigma}\}
\notag\\
   &\qquad\qquad\qquad\qquad{}
   -\frac{1}{3}C_\mu
   (\{\delta A_\nu,\Bar{D}_\rho\delta_\star A_{\star\sigma}\}
   +\{\delta_\star A_{\star\nu},\Bar{D}_\rho\delta A_\sigma\})
   +\frac{1}{3}\partial_\mu
   (C_\nu[\delta A_\rho,\delta_\star A_{\star\sigma}])\biggr].
\label{eq:(2.55)}
\end{align}
Naturally, this parity-odd part is controlled by the gauge anomaly; it can be
confirmed that this combination vanishes when the gauge representation of the
fermion is anomaly free.

For the parity-even part~\eqref{eq:(2.39)}, we have Lorentz symmetry-violating
terms as well as Lorentz-preserving terms. For the latter, we have
(again omitting the symbol~$\tr$),
\begin{align}
   &\left.
   \mathcal{L}(A,A_\star;\delta A,\delta_\star A_\star)
   \right|_{\text{parity-even, Lorentz-preserving}}
\notag\\
   &=\frac{f_0}{a^2}\delta A_\mu\delta_\star A_{\star\mu}
\notag\\
   &\qquad{}
   +\left(-\frac{3f_1}{2}+\frac{f_2}{2}-\frac{f_3}{2}\right)
   [(\Bar{D}_\mu\delta A_\mu) C_\nu\delta_\star A_{\star\nu}
   -C_\mu\delta A_\mu(\Bar{D}_\nu\delta_\star A_{\star\nu})]
\notag\\
   &\qquad{}
   -\left(\frac{f_1}{2}+\frac{f_2}{2}-\frac{3f_3}{2}\right)
   [C_\mu(\Bar{D}_\nu\delta A_\mu)\delta_\star A_{\star\nu}
   -\delta A_\mu C_\nu(\Bar{D}_\mu\delta_\star A_{\star\nu})]
\notag\\
   &\qquad{}
   -\left(\frac{f_1}{2}+\frac{f_2}{2}-\frac{3f_3}{2}\right)
   [C_\nu\delta A_\mu(\Bar{D}_\mu\delta_\star A_{\star\nu})
   -(\Bar{D}_\nu\delta A_\mu)C_\mu\delta_\star A_{\star\nu}]
\notag\\
   &\qquad{}
   +\left(-\frac{7f_1}{2}+\frac{f_2}{2}+\frac{f_3}{2}\right)
   [(\Bar{D}_\mu C_\mu)\delta A_\nu\delta_\star A_{\star\nu}
   -\delta A_\nu(\Bar{D}_\mu C_\mu)\delta_\star A_{\star\nu}]
\notag\\
   &\qquad{}
   -\left(\frac{3f_1}{2}-\frac{f_2}{2}+\frac{f_3}{2}\right)
   [\delta A_\mu C_\mu(\Bar{D}_\nu\delta_\star A_{\star\nu})
   -C_\nu(\Bar{D}_\mu\delta A_\mu)\delta_\star A_{\star\nu}]
\notag\\
   &\qquad{}
   +\left(13f_1-3f_2-3f_3\right)
   (\Bar{D}_\mu\delta A_\mu)(\Bar{D}_\nu\delta_\star A_{\star\nu})
\notag\\
   &\qquad{}
   +\left(9f_1-3f_2-f_3\right)
   (\Bar{D}_\mu\delta A_\nu)(\Bar{D}_\mu\delta_\star A_{\star\nu})
\notag\\
   &\qquad{}
   +\left(-19f_1+5f_2+5f_3\right)
   (\Bar{D}_\nu\delta A_\mu)(\Bar{D}_\mu\delta_\star A_{\star\nu})
\notag\\
   &\qquad{}
   +\left(\frac{11f_1}{6}-\frac{f_2}{6}-\frac{7f_3}{6}\right)
   C_\mu\delta A_\nu C_\mu\delta_\star A_{\star\nu}
\notag\\
   &\qquad{}
   +\left(-\frac{13f_1}{6}+\frac{11f_2}{6}-\frac{7f_3}{6}\right)
   (C_\mu\delta A_\mu C_\nu\delta_\star A_{\star\nu}
   +C_\nu\delta A_\mu C_\mu\delta_\star A_{\star\nu})
\notag\\
   &\qquad{}
   +\left(-\frac{5f_1}{12}+\frac{19f_2}{12}-\frac{17f_3}{12}\right)
   (C_\nu C_\mu \delta A_\mu\delta_\star A_{\star\nu}
   +\delta A_\mu C_\mu C_\nu\delta_\star A_{\star\nu})
\notag\\
   &\qquad{}
   +\left(\frac{19f_1}{12}-\frac{5f_2}{12}-\frac{5f_3}{12}\right)
   (C_\mu C_\nu\delta A_\mu\delta_\star A_{\star\nu}
   +\delta A_\mu C_\nu C_\mu\delta_\star A_{\star\nu})
\notag\\
   &\qquad{}
   +\left(-\frac{17f_1}{12}+\frac{19f_2}{12}-\frac{11f_3}{12}\right)
   (C_\mu C_\mu\delta A_\nu\delta_\star A_{\star\nu}
   +\delta A_\nu C_\mu C_\mu\delta_\star A_{\star\nu})
\notag\\
   &\qquad{}
   +\left(\frac{9f_1}{2}-\frac{3f_2}{2}-\frac{f_3}{2}\right)
   [\partial_\mu(C_\mu\delta A_\nu\delta_\star A_{\star\nu})
   - \partial_\mu(\delta A_\nu C_\mu\delta_\star A_{\star\nu})]
\notag\\
   &\qquad{}
   +\left(f_2-\frac{3f_3}{2}\right)
   [\partial_\mu(C_\nu\delta A_\mu\delta_\star A_{\star\nu})
   -\partial_\nu(\delta A_\mu C_\mu\delta_\star A_{\star\nu})]
\notag\\
   &\qquad{}
   -\left(4f_1-f_2-\frac{f_3}{2}\right)
   [\partial_\nu(C_\mu\delta A_\mu\delta_\star A_{\star\nu})
   -\partial_\mu(\delta A_\mu C_\nu\delta_\star A_{\star\nu})]
\notag\\
   &\qquad{}
   +(10f_1-2f_2-4f_3)
   \{\partial_\nu[\delta A_\mu(\Bar{D}_\mu\delta_\star A_{\star\nu})]
   +\partial_\mu[(\Bar{D}_\nu\delta A_\mu)\delta_\star A_{\star\nu}]\}
\notag\\
   &\qquad{}
   +(-8f_1+2f_2+f_3)
   \{\partial_\mu[\delta A_\mu(\Bar{D}_\nu\delta_\star A_{\star\nu})]
   +\partial_\nu[(\Bar{D}_\mu\delta A_\mu)\delta_\star A_{\star\nu}]
\notag\\
   &\qquad\qquad\qquad\qquad\qquad\qquad{}
   +\partial_\mu[\delta A_\nu(\Bar{D}_\mu\delta_\star A_{\star\nu})]
   +\partial_\mu[(\Bar{D}_\mu\delta A_\nu)\delta_\star A_{\star\nu}]\}.
\label{eq:(2.56)}
\end{align}

For the parity-even, Lorentz-violating part,
\begin{align}
   &\left.
   \mathcal{L}(A,A_\star;\delta A,\delta_\star A_\star)
   \right|_{\text{parity-even, Lorentz-violating}}
\notag\\
   &=\frac{3}{2}\left(9f_1-f_2-f_3-\frac{f_4}{2}-\frac{f_5}{2}\right)
   [(\Bar{D}_\nu C_\nu)\delta A_\nu\delta_\star A_{\star\nu}
   -\delta A_\nu(\Bar{D}_\nu C_\nu)\delta_\star A_{\star\nu}]
\notag\\
   &\qquad{}
   -\left(9f_1-f_2-f_3-\frac{f_4}{2}-\frac{f_5}{2}\right)
   (\Bar{D}_\nu\delta A_\nu)(\Bar{D}_\nu\delta_\star A_{\star\nu})
\notag\\
   &\qquad{}
   +\left(\frac{47f_1}{2}-\frac{7f_2}{2}-\frac{7f_3}{2}
   +\frac{f_4}{4}-\frac{7f_5}{4}\right)
   C_\nu\delta A_\nu C_\nu\delta_\star A_{\star\nu}
\notag\\
   &\qquad{}
   +\left(\frac{67f_1}{4}-\frac{11f_2}{4}-\frac{11f_3}{4}
   +\frac{5f_4}{8}-\frac{11f_5}{8}\right)
   (C_\nu C_\nu\delta A_\nu\delta_\star A_{\star\nu}
   +\delta A_\nu C_\nu C_\nu\delta_\star A_{\star\nu})
\notag\\
   &\qquad{}
   +\frac{1}{2}\left(9f_1-f_2-f_3-\frac{f_4}{2}-\frac{f_5}{2}\right)
   [\partial_\nu(\delta A_\nu C_\nu\delta_\star A_{\star\nu})
   -\partial_\nu(C_\nu\delta A_\nu\delta_\star A_{\star\nu})]
\notag\\
   &\qquad{}
   +2\left(9f_1-f_2-f_3-\frac{f_4}{2}-\frac{f_5}{2}\right)
   \{\partial_\nu[\delta A_\nu(\Bar{D}_\nu\delta_\star A_{\star\nu})]
   +\partial_\nu[(\Bar{D}_\nu\delta A_\nu)\delta_\star A_{\star\nu}]\}.
\label{eq:(2.57)}
\end{align}

The local functional $\mathcal{L}(A,A_\star;\delta A,\delta A_\star)$
in~Eq.~\eqref{eq:(1.3)} is given by the sum of~Eqs.~\eqref{eq:(2.55)},
\eqref{eq:(2.56)}, and~\eqref{eq:(2.57)}. In particular,
$\mathcal{L}(A,A_\star=0;\delta A,\delta_\star^\omega A_\star|_{A_\star=0})$
in~Eq.~\eqref{eq:(1.12)} is given by setting $A_{\star\mu}(x)=0$
and~$\delta_\star A_{\star\mu}(x)=\partial_\mu\omega(x)$
(and thus $C_\mu=-A_\mu$, $\Bar{A}_\mu=(1/2)A_\mu$,
and~$\Bar{D}_\mu=\partial_\mu+(1/2)[A_\mu,\cdot]$) in the above expressions.
We see that
$\mathcal{L}(A,A_\star=0;\delta A,\delta_\star^\omega A_\star|_{A_\star=0})$ does
not vanish even if the gauge representation is anomaly free. For example,
from the first term of~Eq.~\eqref{eq:(2.56)},
\begin{equation}
   \mathcal{L}(A,A_\star=0;\delta A,\delta_\star^\omega A_\star|_{A_\star=0})
   =\frac{f_0}{a^2}\tr\delta A_\mu(x)\partial_\mu\omega(x)+\dotsb,
\label{eq:(2.58)}
\end{equation}
and the relation~\eqref{eq:(1.12)} tells us that this corresponds to the mass
term of the gauge field, $(f_0/2a^2)\tr A_\mu(x)A_\mu(x)$, in the effective
action~$\ln\mathcal{Z}[A,0]$. Other terms in~Eqs.~\eqref{eq:(2.56)}
and~\eqref{eq:(2.57)} can be understood in a similar manner, as shown
in~Appendix~\ref{sec:A}. Such gauge-breaking terms can always be removed by
local counterterms (see Appendix~\ref{sec:A}), but such a necessity for
counterterms for restoring the gauge symmetry will be undesirable from the
perspective of a non-perturbative formulation of chiral gauge theories.

\section*{Acknowledgments}
We are grateful to Shoji Hashimoto, Yoshio Kikukawa, and Ken-ichi Okumura for
valuable remarks.
We would like to thank Ryuichiro Kitano and Katsumasa Nakayama for
intensive discussions on a related subject.
The work of H.~S. is supported in part by JSPS Grant-in-Aid for Scientific
Research Grant Number~JP16H03982.

\appendix

\section{Computation of $\delta^\omega\ln\mathcal{Z}[A,0]$}
\label{sec:A}
In this appendix, we obtain the explicit form of the ``gauge
symmetry-breaking'' $\delta^\omega\ln\mathcal{Z}[A,0]$ in~Eq.~\eqref{eq:(1.12)}
and confirm that the breaking can be removed by local counterterms when the
gauge representation is anomaly free. We note the identity
\begin{align}
   \delta^\omega\ln\mathcal{Z}[A,0]
   &=\int_0^1d\xi\,\frac{d}{d\xi}\left(\delta^\omega\ln\mathcal{Z}[A,0]
   \right)_{A\to \xi A}
\notag\\
   &=\int_0^1d\xi\,\left(\delta\delta^\omega\ln\mathcal{Z}[A,0]
   \right)_{A\to \xi A,\delta A\to A}
\notag\\
   &=\int d^4x\,
   \int_0^1d\xi\,\mathcal{L}
   (\xi A,A_\star=0;A,\delta_\star^\omega A_\star|_{A_\star=0}),
\label{eq:(A1)}
\end{align}
where we have assumed $\left(\delta^\omega\ln\mathcal{Z}[A,0]\right)_{A=0}=0$.
Note that $\delta_\star^\omega A_{\star\mu}|_{A_\star=0}=\partial_\mu\omega$. Then
using Eqs.~\eqref{eq:(2.55)}, \eqref{eq:(2.56)}, and~\eqref{eq:(2.57)} in the
above formula, we find (omitting the symbol~$\int d^4x\,\tr$)
\begin{equation}
   \left.\delta^\omega\ln\mathcal{Z}[A,0]\right|_{\text{parity-odd}}
   =\frac{1}{24\pi^2}\epsilon_{\mu\nu\rho\sigma}
   (\partial_\mu\omega)\left(A_\nu\partial_\rho A_\sigma
   +\frac{1}{2}A_\nu A_\rho A_\sigma\right),
\label{eq:(A2)}
\end{equation}
\begin{align}
   &\left.\delta^\omega\ln\mathcal{Z}[A,0]
   \right|_{\text{parity-even, Lorentz-preserving}}
\notag\\
   &=\frac{f_0}{a^2}(\partial_\mu\omega)A_\mu
\notag\\
   &\qquad{}
   +(-13f_1+3f_2+3f_3)(\partial_\mu\omega)\partial_\mu\partial_\nu A_\nu
   +(10f_1-2f_2-4f_3)(\partial_\mu\omega)\partial_\nu\partial_\nu A_\mu
\notag\\
   &\qquad\qquad{}
   +(-5f_1+f_2+2f_3)(\partial_\mu\omega)[\partial_\mu A_\nu,A_\nu]
   +(8f_1-2f_2-f_3)(\partial_\mu\omega)[\partial_\nu A_\nu,A_\mu]
\notag\\
   &\qquad\qquad\qquad{}
   +\frac{1}{3}
   (-11f_1+7f_2-2f_3)(\partial_\mu\omega)
   (A_\mu A_\nu A_\nu+A_\nu A_\nu A_\mu)
\notag\\
   &\qquad\qquad\qquad\qquad{}
   +\frac{1}{3}
   (19f_1-5f_2-5f_3)(\partial_\mu\omega)A_\nu A_\mu A_\nu,
\label{eq:(A3)}
\end{align}
and
\begin{align}
   &\left.\delta^\omega\ln\mathcal{Z}[A,0]
   \right|_{\text{parity-even, Lorentz-violating}}
\notag\\
   &=\left(9f_1-f_2-f_3-\frac{f_4}{2}-\frac{f_5}{2}\right)
   (\partial_\mu\omega)
   \left(\partial_\mu\partial_\mu A_\mu
   -[\partial_\mu A_\mu,A_\mu]
   \right)
\notag\\
   &\qquad{}
   +\frac{1}{3}
   \left(57f_1-9f_2-9f_3+\frac{3f_4}{2}-\frac{9f_5}{2}\right)
   (\partial_\mu\omega)A_\mu A_\mu A_\mu.
\label{eq:(A4)}
\end{align}

The parity-odd breaking term~\eqref{eq:(A2)} is, as expected, the consistent
gauge anomaly associated with a single left-handed Weyl fermion. This cannot be
written as the gauge variation of a local term and vanishes if the gauge
representation is anomaly free.

Concerning the parity-even breaking terms~\eqref{eq:(A3)} and~\eqref{eq:(A4)},
it must be possible to rewrite them as the gauge variation of local terms. In
fact, we can see that (again omitting the symbol~$\int d^4x\,\tr$)
\begin{align}
   &\left.\delta^\omega\ln\mathcal{Z}[A,0]
   \right|_{\text{parity-even, Lorentz-preserving}}
\notag\\
   &=\delta^\omega
   \biggl[
   \frac{f_0}{2a^2}A_\mu A_\mu
\notag\\
   &\qquad\qquad{}
   +\frac{1}{2}
   (-13f_1+3f_2+3f_3)A_\mu\partial_\mu\partial_\nu A_\nu
\notag\\
   &\qquad\qquad\qquad{}
   +(5f_1-f_2-2f_3)
   (A_\mu\partial_\nu\partial_\nu A_\mu
   -A_\mu A_\nu\partial_\mu A_\nu
   +A_\mu A_\nu\partial_\nu A_\mu)
\notag\\
   &\qquad\qquad\qquad\qquad{}
   +\frac{2}{3}
   (f_1+f_2-2f_3)A_\mu A_\mu A_\nu A_\nu
   +\frac{1}{12}
   (-11f_1+f_2+7f_3)A_\mu A_\nu A_\mu A_\nu
   \biggr],
\label{eq:(A5)}
\end{align}
and
\begin{align}
   &\left.\delta^\omega\ln\mathcal{Z}[A,0]
   \right|_{\text{parity-even, Lorentz-preserving}}
\notag\\
   &=\delta^\omega
   \biggl[
   \frac{1}{2}
   \left(9f_1-f_2-f_3-\frac{f_4}{2}-\frac{f_5}{2}\right)
   A_\mu\partial_\mu\partial_\mu A_\mu
\notag\\
   &\qquad\qquad{}
   +\frac{1}{12}
   \left(57f_1-9f_2-9f_3+\frac{3f_4}{2}-\frac{9f_5}{2}\right)
   A_\mu A_\mu A_\mu A_\mu
   \biggr].
\label{eq:(A6)}
\end{align}
Thus, these breakings can be removed by local counterterms.

\end{document}